\documentstyle[prd,aps,floats]{revtex}
\begin{document}
\vbox{
\halign{&#\hfill\cr
        & MADPH-96-953 \cr
        & UTTG-11-96 \cr
        & McGill-96-26 \cr
        & UdeM-GPP-TH-96-40 \cr} }

\bigskip
\bigskip
\bigskip
\bigskip

\centerline{{ \LARGE\bf NRQCD matrix elements in polarization}}
\bigskip
\centerline{{ \LARGE\bf  of J/Psi produced from b-decay}}
\bigskip
\bigskip
\bigskip

\centerline{{\large Sean Fleming}
             \footnote{Email: fleming@phenxs.physics.wisc.edu }}
\smallskip
\centerline{{\it Department of Physics, University of Wisconsin, Madison, WI
53706}}
\bigskip
\centerline{{\large Oscar F. Hern\'andez}
             \footnote{Email: oscarh@hep.physics.mcgill.ca}}
\smallskip
\centerline{{\it Laboratoire de Physique Nucl\'eaire,
Universit\'e de Montr\'eal, C.P. 6128,
Succ. A, Montr\'eal, Canada H3C 3P8}}
\bigskip
\centerline{{\large Ivan Maksymyk}
             \footnote{Email: maksymyk@physics.utexas.edu}}
\smallskip
\centerline{{\it Theory Group, Department of Physics, University of Texas,
         Austin, TX 78712}}
\bigskip
\centerline{{\large H\'el\`ene Nadeau}
             \footnote{Email: nadeau@hep.physics.mcgill.ca}}
\smallskip
\centerline{{\it Department of Physics, McGill University, Montr\'eal, Canada
H3A 2T8}}
\bigskip

\begin{abstract}
We present the non-relativistic QCD (NRQCD) prediction for the
polarization of the $J/\psi$ produced in $b \to J/\psi +X$, as well as
the helicity-summed production rate. We propose that these observables
provide a means of measuring the three most important color-octet
NRQCD matrix elements involved in $J/\psi$ production.
Anticipating the measurement of the polarization parameter
$\alpha$, we determine
its expected range given
current experimental bounds on the color-octet matrix elements.
\end{abstract}

\section{ Introduction}

A rigorous theoretical framework within which quarkonium production can
be studied is provided by the nonrelativistic QCD (NRQCD) factorization
formalism developed by Bodwin, Braaten, and Lepage~\cite{bbl}. This
approach is based on NRQCD~\cite{casslep}, an effective field theory that
can be made equivalent
to full QCD to any desired order in heavy-quark relative
velocity. The NRQCD
factorization formalism is not a model, but rather a
rigorous derivation
within NRQCD of a
factorized form for quarkonium production and decay rates.

A central result of the NRQCD factorization formalism is that inclusive
quarkonium production cross sections must have the form of a sum of
products of short-distance coefficients and NRQCD matrix elements. The
short-distance coefficients are associated with the production of a heavy
quark-antiquark pair in a specific color and angular-momentum state. They
can be calculated using ordinary perturbative techniques. As to the NRQCD
matrix elements, they parameterize the effect of long-distance physics
such as the hadronization of the quark-antiquark pair. These can
be determined phenomenologically.

The power of the NRQCD formalism stems {}from the fact that
factorization formulas for observables
are essentially expansions in the small parameter
$v^2$, where $v$ is the average
relative velocity of the heavy quark and anti-quark in
the quarkonium bound state. $v^2\sim 0.3$ for charmonium,
and 0.1 for bottomonium.  NRQCD
$v$-scaling rules~\cite{lmnm}
allow us to estimate the relative sizes of various NRQCD
matrix elements.  This information, along with
the dependence
of the short-distance coefficients on coupling constants,
permits us to decide which terms must be retained in
expressions for observables
so as to reach a given level of accuracy.
Generally, to leading order, factorization formulas involve only a few
matrix elements, so several observables can be related by a small number
of parameters.

NRQCD predictions therefore presuppose QCD, factorization, and a
reasonably
convergent $v^2$ expansion.  If the quantitative predictions of
NRQCD fail, one of these three assumptions is failing,
most likely the validity of the $v^2$ expansion.

In many instances of $J/\psi$ production,
the most important NRQCD matrix elements are
$\langle{\cal O}^\psi_1({}^3S_1) \rangle$,
$\langle{\cal O}^\psi_8({}^3S_1) \rangle$,
$\langle{\cal O}^\psi_8({}^1S_0) \rangle$, and
$\langle{\cal O}^\psi_8({}^3P_J) \rangle$. These four quantities
parametrize the hadronization into a $J/\psi$ boundstate of a
$c\bar{c}$ pair produced initially with the stated quantum numbers (angular
momentum ${}^{2S+1}L_J$, color quantum number 1 or 8).
Previous to the development
of the factorization formalism of Ref.~\cite{bbl}, most
$J/\psi$ production calculations took into account only the hadronization of
$c\bar{c}$ pairs initially produced in the color-singlet ${}^3S_1$
state, as parameterized by
$\langle{\cal O}^\psi_1({}^3S_1) \rangle$.

Recently, in regard to the three most important
color-octet matrix elements involved in
$J/\psi$ production, it has been noticed that there prevails a
certain inconsistency between the values
measured at CDF \cite{cl} and those measured in other processes
\cite{toohigh}.  We propose here
that inclusive
$J/\psi$ production {}from $b$-decay can be of avail in unravelling
this difficulty by serving as a supplementary arena for measuring
the contentious matrix elements.

Inclusive production of $J/\psi$ {}from $b$-decay
provides two measurable combinations of the matrix elements. The first
one is the helicity-summed rate
$\Gamma(b \to J/\psi + X)$.
The second combination concerns the polarization parameter
$\alpha$ appearing in the electromagnetic decay rate of $J/\psi$ to
lepton pairs:
\begin{equation}
\label{thetadefined}
\frac{d\Gamma}{dcos\theta}\big( \psi \rightarrow \mu^+\mu^-(\theta)
\big)
\propto 1 + \alpha \cos^2{\theta} \; ,
\end{equation}
where the polar angle $\theta$ is defined in the $J/\psi$ rest frame for
which the $z$-axis is aligned with the direction of motion of the
$J/\psi$ in the lab.

The short-distance physics for the process
$b\to J/\psi + X$
is described by the four-quark
Fermi interactions $b \rightarrow c\bar{c}s$ and
$b \rightarrow c\bar{c}d$,  at the $m_b$ scale.
The Feynman diagram is shown in Fig.~1.
Let us define $P$ to be the total
four-momentum of the $c\bar{c}$ system.
The $s$ and $d$ quarks are assumed
massless.

\section{Factorization formula for the spin-summed production rate}

We first calculate the factorization formula for the
spin-summed $J/\psi$ production rate.
Let us define ${\bf q}$ as the relative
momentum of the $c$ and $\bar{c}$ quarks evaluated in the rest frame
of the $c\bar{c}$ pair.  The NRQCD matching formalism requires only that we
know
the algebraic form of the amplitude for small values of ${\bf q}$.
This is because the hadronization of the heavy quark pair into
quarkonium occurs with nonnegligible probability only when
$|{\bf q}| \ll m_c$.  Moreover, this is the regime
in which the NRQCD effective lagrangian is valid.
Therefore, we expand the Feynman amplitude
in powers of ${\bf q}$ and keep only pieces up to linear order.
The amplitude
for $b\to c\bar{c}s$ is given by
\begin{eqnarray}
\label{bdecayfeynamp-reduced}
{\cal M}(\sigma,\tau;c,d,e,f;{\bf q}) \;\; & = & \;\;
\frac{G_F}{\sqrt{2}}\; V_{cb}V_{cs}^*
\left[
\frac{(2C_+ - C_-)}{3}
\delta^{cd}\delta^{ef} \;
+ \;
(C_+ + C_-) T^g_{cd}T^g_{ef}\right]
\;\;
\bar{u}_s\gamma^\mu ( 1 - \gamma_5 ) u_b
\nonumber\\
& \times &
\bigg[ 2m_c \; \; \Lambda^\mu_i \xi_\sigma^\dagger \! \sigma^i \!
\eta_\tau \; - \;
P^\mu \;\;\xi_\sigma^\dagger \! \eta_\tau  \; + \;
2i \Lambda^\mu_k\epsilon^{mik} \;\;\xi_\sigma^\dagger \! q^m \! \sigma^i \!
\eta_\tau
\bigg] \bigg( 1 + O\left({ {\bf q}^2 \over m^2_c }\right) \bigg) \; ,
\end{eqnarray}
where $\xi$ and $\eta$ are non-relativistic two-spinors for the $c$
and $\bar{c}$ quarks respectively; $c,d,e,f$ are color indices and
$\sigma$ and $\tau$ are the individual spins of the charm quarks.  The
factors involving $C_+$ and $C_-$ are Wilson coefficients which govern
the scale-evolution of the four-quark Fermi interaction.
$\Lambda^\mu_i$ (defined in detail in Ref.~\cite{bc}) is the Lorentz boost
matrix that takes a three-vector {}from the $c\bar{c}$ rest frame to the
lab frame . In the above equation, the term with the color
Kr\"onecker deltas contributes to the production of $c\bar{c}$ pairs in
color-singlet states, and the term with color matrices $T^g$ contributes
to the production of $c\bar{c}$ pairs in color-octet states.

{}From Eq.~\ref{bdecayfeynamp-reduced}, using the matching procedure
presented in Ref~\cite{makflem}, we determine the spin-summed
rate for $b \to J/\psi + X$ in the NRQCD factorization formalism:
\begin{eqnarray}
\label{totalfacform}
\Gamma(b \rightarrow J/\psi + X) & = &
\frac{G_F^2}{864\pi}\; \frac{(m_b^2 - 4 m_c^2)^2}{m_b^3 m_c}
\Big| V_{cb} \Big|^2 \;\;
\Bigg( 2(2C_+ - C_-)^2 (m_b^2 + 8 m_c^2)
         \big\langle{\cal O}^\psi_1({}^3S_1) \big\rangle \nonumber\\
& + & \;\;
       3(C_+ + C_-)^2(m_b^2 + 8 m_c^2)
       \big\langle{\cal O}^\psi_8({}^3S_1) \big\rangle \;\; + \;\;
 9(C_+ + C_-)^2 m_b^2
\big\langle{\cal O}^\psi_8({}^1S_0) \big\rangle \nonumber\\
& + & \;\;
  6(C_+ + C_-)^2 (m_b^2 + 8 m_c^2)
  \frac{\big\langle{\cal O}^\psi_8({}^3P_1) \big\rangle}{m_c^2} \;\;
\Bigg) \Big( 1 + O(v^2) \Big) \;
\end{eqnarray}
where, upon summing
over the light quarks $s$ and $d$, we
have used $|V_{cb}V_{cs}^*|^2 + |V_{cb}V_{cd}^*|^2
\approx |V_{cb}|^2$. We have taken the inital $b$-quark to be
unpolarized.  Here
$\langle{\cal O}^{\psi}_n\rangle\equiv\langle 0|{\cal O}^{\psi}_n|0\rangle$
are NRQCD $J/\psi$ production matrix elements.
We consider only the leading color-singlet piece and the leading
color-octet pieces in the relativistic $v^2$-expansion.  Our
calculation of the coefficients in the above spin-summed factorization
formula concurs with the results presented in Ref.~\cite{kls}.

According to the NRQCD $v$-scaling rules, the color-octet matrix
elements in Eq.~\ref{totalfacform}
are all expected to be suppressed by $v^4$
with respect to the basic color-singlet matrix element
$\langle{\cal O}^\psi_1({}^3S_1) \rangle$.
Thus, at the outset,
one would not expect the octet contributions to play a major role
in the production of $J/\psi$ articles.
However, as pointed out in Ref.~\cite{prolett},
the color-singlet coefficient $2C_+ - C_-$ decreases
significantly as it is evolved down {}from its value of 1
at the scale $M_W$ to its value of roughly 0.40 at the scale
$m_b$.  On the other hand, the color-octet coefficient
$(C_+ + C_-)/2$ increases slightly {}from 1 at the scale $M_W$
to roughly 1.10 at the scale $m_b$.
For this reason,
the octet contributions are actually
as important as, or more important than, the basic singlet
contribution.
In fact, the short-distance coefficients
in the color-octet terms are
some 50 times larger than those in the singlet term!

\section{Factorization formula for helicity rates}

In the previous section, we discussed how the serendipitous Wilson-coefficient
enhancement of the octet contributions
enables us to measure the color-octet matrix elements contributing to
the helicity-summed rate
$\Gamma( b\rightarrow J/\psi + X) $. This enhancement can be further
exploited to determine the color-octet matrix elements
by considering the polarization of the produced
$J/\psi$'s, as measured by the parameter $\alpha$
in Eq.~\ref{thetadefined} . It can be calculated via the formula
\begin{equation}
\label{alphasig}
\alpha =  \frac{\sigma(+) + \sigma(-) - 2 \sigma(0)}
                       {\sigma(+) + \sigma(-) + 2 \sigma(0)}
\end{equation}
where $\sigma(\lambda)$ represents the production rate
of $J/\psi$ with helicity $\lambda$.

Since $\alpha$ is a ratio, we need only calculate the relative sizes
of the various contributions to the helicity rates.  Thus, we need
only calculate the effective Feynman amplitude squared, evaluated in
the $b$ rest frame, with the $J/\psi$ momentum ${\bf P}$ oriented in
the positive $z$-direction.  In this case, the angular momentum of the
$J/\psi$ in the canonical $z$-direction corresponds simply to the
particle's helicity.

Beneke and Rothstein \cite{br} and Braaten and Chen \cite{bc} have
developed techniques for deriving the production rates of
quarkonia with specified helicities.  Applying
their
methods to the amplitude given in
Eq.~\ref{bdecayfeynamp-reduced},
we obtain
\begin{eqnarray}
\label{Rlambda}
\Gamma \big( b \to J/\psi(\lambda) + X \big)   & \propto&
2 (2C_+ - C_-)^2
\big\langle{\cal O}^\psi_1({}^3S_1) \big\rangle
\left[ (m_b^2 - 4 m_c^2) \delta_{\lambda 0} + 4 m_c^2 (1 - \lambda) \right]
\nonumber\\
& + & 3 (C_+ +  C_-)^2
\big\langle{\cal O}^\psi_8({}^3S_1) \big\rangle
\left[ (m_b^2 - 4 m_c^2) \delta_{\lambda 0} + 4 m^2 (1 - \lambda) \right]
\nonumber\\
& + & 3 (C_+ +  C_-)^2
\big\langle{\cal O}^\psi_8({}^1S_0) \big\rangle
m_b^2
\nonumber\\
& + & 9 (C_+ +  C_-)^2
\frac{\big\langle{\cal O}^\psi_8({}^3P_0) \big\rangle}{m_c^2}
\left[ (m_b^2 - 4 m_c^2) (1 - \delta_{\lambda 0})
+ 8 m_c^2  - 4 m_c^2 \lambda \right].
\end{eqnarray}
In displaying Eq.~\ref{Rlambda},
we have adopted the standard practice of rewriting the
matrix element $\langle {\cal O}_8({}^3P_1) \rangle$ using
the relation
\begin{equation}
\label{simplification}
\big\langle{\cal O}^\psi_8({}^3P_J) \big\rangle
\simeq
(2J + 1)
\big\langle{\cal O}^\psi_8({}^3P_0) \big\rangle .
\end{equation}
This latter relation is due to heavy-quark spin symmetry, and is
valid up to relative order $v^2$.
It must be kept in mind, however, that $b$-decay does not
actually produce $c\bar{c}$ states in the ${}^3P_0$ configuration
at leading order in the coupling constants.

Some consequences of angular momentum conservation and of the
left-handedness of the charged-current Fermi interaction
are illustrated in Fig.~2.  The thin arrow labelled
${\bf \stackrel{\rightarrow}{P} }$ represents the trajectory of
the center of mass of the $c\bar{c}$ system.  The other thin arrow
represents the trajectory of the $s$ or $d$ quark.  Thick arrows
denote intrinsic angular momenta, with $\lambda_{c\bar{c}}$ denoting
the helicity of the $c\bar{c}$ system
prior to hadronization.  Since the charged-current
Fermi interaction couples only the left-handed parts of the fermion
fields, the masslessness of the $s$ and $d$ quarks
ensures that they are emitted with negative
helicity.  Therefore, due to angular momentum conservation,
the helicity outcome $\lambda_{c\bar{c}} = +1$ is not allowed.  This
has certain obvious consequences for the $J/\psi$ helicity production
rates:  the basic color-singlet contribution (parametrized by
$\langle{\cal O}^\psi_1({}^3S_1) \rangle$ and involving the direct
hadronization of a color-singlet ${}^3S_1$ state into a $J/\psi$)
is zero for $\lambda = +1$, at lowest order in the relativistic expansion.
The same is also true of the $\langle{\cal O}^\psi_8({}^3S_1) \rangle$
contribution, which involves the hadronization of a color-octet
${}^3S_1$ state into a $J/\psi$ via the heavy-quark-spin-preserving
$L=0$ emission or absorption of two soft gluons~\cite{cw}.

The color-singlet contribution to the helicity rates of the $J/\psi$
produced in $b$-decay was calculated in the color-singlet model
by M. Wise in Ref.~\cite{wise} 16 years ago.  This author presented
expressions for the production rates of longitudinal and transverse
helicities, given in terms of the color-singlet radial wave function at the
origin.
The result in Ref.~\cite{wise} concurs with the first line of our
Eq.~\ref{Rlambda}.

\section{Prediction for the polarization parameter alpha}

While experimental determinations of
helicity-summed
$\Gamma(b\to J/\psi + X)$ have already been carried out~\cite{pdb}, a
measurement of the polarization parameter
$\alpha$ is not yet available.
Anticipating the availability of this latter measurement, it is
interesting to determine the range of $\alpha$ which
is predicted by the NRQCD factorization formalism.
Using Eqs.~\ref{alphasig} and \ref{Rlambda}, we can express $\alpha$
in terms of the NRQCD matrix elements.
Before proceding, however, we must first decide
which value of $m_b$ to use.
Since we are presenting a leading-order calculation, we
may in principle choose to
use either the pole mass or the $\overline{\mbox{MS}}$
mass for $m_b$, the difference between these two being merely a
higher-order-in-$\alpha_s$ effect. However,
it turns out that the
leading-order NRQCD prediction of
the polarization parameter $\alpha$ depends quite strongly on the
choice of $m_b$.
Therefore we report
results for a wide range of $m_b$, {}from 4.1~GeV to 5.3~GeV. This range
includes the values of $m_b=4.3\pm0.2$ GeV corresponding to
the $\overline{\mbox{MS}}$ mass, and the values $m_b=5.0\pm0.2$ GeV
corresponding to the pole mass determined on the lattice
\cite{bodwin}. Our range
is centered around $m_b = 4.7$~GeV, which is the central value
for the pole mass~\cite{pdb}. This choice is consistent with the fact
that the numerical value that we use
for $\langle{\cal O}^\psi_1({}^3S_1)\rangle$ is taken {}from
Ref.~\cite{bodwin}, in which the pole mass
was used in the extraction of the matrix element value.

For concreteness, we present in the text
below
only
our results for
the choice $m_b = 4.7$ GeV.  Using $C_+(m_b) = 0.868$,
$C_-(m_b) =1.329$ and $ m_c = 1.55 $ GeV, one
obtains the leading-order NRQCD prediction for the polarization
parameter
\begin{equation}
\label{bigresult}
\alpha = \frac{
         -0.39 \big\langle{\cal O}^\psi_1({}^3S_1) \big\rangle
     -   17    \big\langle{\cal O}^\psi_8({}^3S_1) \big\rangle
     +   52    \big\langle{\cal O}^\psi_8({}^3P_0) \big\rangle / m^2_c
}
{
              \big\langle{\cal O}^\psi_1({}^3S_1) \big\rangle
     +  44    \big\langle{\cal O}^\psi_8({}^3S_1) \big\rangle
     +  61    \big\langle{\cal O}^\psi_8({}^1S_0) \big\rangle
     + 211    \big\langle{\cal O}^\psi_8({}^3P_0) \big\rangle / m^2_c
} \;  .
\end{equation}

It must be noted that $\alpha$ depends only weakly on
$\langle{\cal O}^\psi_8({}^3S_1) \rangle$
and most strongly on
$\langle{\cal O}^\psi_8({}^3P_0) \rangle / m^2_c$.

We now procede to determine the range of $\alpha$
which is
consistent with existing information on the matrix elements
$\langle{\cal O}^\psi_1({}^3S_1) \rangle$,
$\langle{\cal O}^\psi_8({}^3S_1) \rangle$,
$\langle{\cal O}^\psi_8({}^1S_0) \rangle$, and
$\langle{\cal O}^\psi_8({}^3P_0) \rangle/m_c^2$
and with various constraints on linear combinations of these quantities.
We first review this information:
\begin{itemize}

\item Bodwin et al.~\cite{bodwin} have determined the color-singlet
matrix element $\langle{\cal O}^\psi_1({}^3S_1) \rangle$  {}from decay
rates of $J/\psi$ to lepton pairs to be $\langle{\cal
O}^\psi_1({}^3S_1) \rangle = 1.1 \pm 0.1 \; \mbox{GeV}^3$. The error
reflects theoretical uncertainties due to higher order $\alpha_s$ and
$v^2$ corrections. This phenomenologically extracted value for the
color-singlet matrix element is in agreement with the lattice
calculation result also
presented in~Ref.~\cite{bodwin}.

\item A constraint on the octet matrix elements is
given by the requirement that the theoretical prediction for the
spin-summed production rate $\Gamma(b \to J/\psi + X)$ given in
Eq.~\ref{totalfacform} be consistent with the experimentally measured
value of $3.42 \, \mu{\rm eV}$.  This constraint can be expressed as
\begin{equation}
\label{decayconstraint}
0.342~{\rm GeV}^3 = 0.096 \big\langle{\cal O}^{\psi}_1(^3S_1) \big\rangle
        + 4.21 \big\langle{\cal O}^\psi_8({}^3S_1) \big\rangle
        + 6.76\big\langle{\cal O}^\psi_8({}^1S_0) \big\rangle
        + 25.3  {\big\langle{\cal O}^\psi_8({}^3P_0) \big\rangle
        \over m^2_c}  \; .
\end{equation}
To compute the above experimental value of $\Gamma(b \to J/\psi + X) =
3.42\mu\, $eV, we have used
$BR\big(B({\rm
charge~not~determined})\to J/\psi({\rm direct})+X\big)$
=(0.80$\pm$0.08)\%, and $\tau$(averaged~over~$B$~hadrons)=1.54~picoseconds.
To compute the short-distance coefficients on the right-hand-side
of Eq.~\ref{decayconstraint}, we have taken
$G_F=1.1664 \times 10^{-5}$ GeV$^{-2}$~\cite{pdb} and
$|V_{cb}|=0.0381\pm0.0021$~\cite{beautyrome}.
One can attribute an error of roughly 30\% to the above equation, to
reflect relativistic corrections and uncertainties in the scale used
to evaluate the Wilson coefficients~\cite{prolett,be}.
It must be
pointed out that the color-singlet model prediction
for
spin-summed $J/\psi$ production
 --- obtained by setting
all color-octet matrix elements to zero, and using  $\langle{\cal
O}^\psi_1({}^3S_1) \rangle = 1.1 \pm 0.1 \; \mbox{GeV}^3$ --- is
roughly
one-third of the experimental value. However the NRQCD
prediction for the $b$-decay width
can be made to agree
with the experimental
measurements for reasonable values of the color-octet matrix elements.

\item Cho and Leibovich~\cite{cl} have performed a fit
to CDF data over low and high $p_T$ ranges and have found
\begin{equation}
\label{clmconstraint}
\big\langle{\cal O}^\psi_8({}^3S_1) \big\rangle  =  0.0066 \pm
0.0021~{\rm GeV}^3 \;
\end{equation}
and
\begin{equation}
\label{clspconstraint}
\big\langle {\cal O}^{\psi}_8(^1S_0)  \big\rangle + 3
{\big\langle {\cal O}^{\psi}_8(^3P_0)  \big\rangle \over m^2_c}  =
0.066\pm0.015~{\rm GeV}^3\; .
\end{equation}

\item {}From photoproduction data,
Amundson et al.~\cite{afm} have determined
\begin{equation}
\big\langle {\cal O}^{\psi}_8(^1S_0)  \big\rangle + 7
{\big\langle {\cal O}^{\psi}_8(^3P_0)  \big\rangle \over m^2_c}
= 0.020\pm0.001~{\rm GeV}^3 \; .
\label{afmbound}
\end{equation}

Note that the errors quoted in
Eqs.~\ref{clmconstraint},~\ref{clspconstraint}, and~\ref{afmbound} are
statistical only, and that the analyses
in \cite{cl} and \cite{afm}
did not take into account
theoretical uncertainties
due to next-to-leading order (NLO) corrections.
However we expect NLO corrections to be very large (see~\cite{cg}
for a discussion of NLO corrections in hadronic $J/\psi$ production
calculations). Moreover the photoproduction results of
Ref.~\cite{afm} may also have significant higher-twist
corrections~\cite{br}.

\item Finally, due to $v$-scaling arguments, we expect
the color-octet matrix elements to be roughly of order
$v^4\langle{\cal O}^\psi_1({}^3S_1) \rangle$.

\end{itemize}

With this information in mind,
we now determine the range of values of $\alpha$
that we can expect.
Our method of determining this range consists of
``scanning" through the three-dimensional parameter space
of the color-octet matrix elements,
and determining the
maximum and minimum value of $\alpha$ that occur in the
allowed volume.
Our scan is subject to the following constraints:
we allow $\langle{\cal O}^\psi_8({}^3S_1) \rangle$ to vary in the range
\begin{equation}
\label{clm2}
\big\langle{\cal O}^\psi_8({}^3S_1) \big\rangle  \in
[0.003,.014]~{\rm GeV}^3 \;  ;
\end{equation}
we allow the photoproduction-related linear combination
of $\langle {\cal O}^{\psi}_8(^1S_0)\rangle$
and $\langle {\cal O}^{\psi}_8(^3P_0)\rangle$,
to vary
in the range
\begin{equation}
\label{afm2}
\big\langle {\cal O}^{\psi}_8(^1S_0)  \big\rangle + 7
{\big\langle {\cal O}^{\psi}_8(^3P_0)  \big\rangle \over m^2_c} \in
[0.0,0.04]~{\rm GeV^3}\; ,
\end{equation}
where the range we have chosen takes into account theoretical
uncertainties due to NLO corrections and possibly
significant hight-twist corrections;
we allow the b-decay-related linear combination
given in Eq.~\ref{decayconstraint} to vary $\pm 30\%$ about the
central value of $0.342\, {\rm GeV}^3$, {\it i.e.}
\begin{equation}
\label{decayconstraint2}
0.096 \big\langle{\cal O}^{\psi}_1(^3S_1) \big\rangle
        + 4.21 \big\langle{\cal O}^\psi_8({}^3S_1) \big\rangle
        + 6.76\big\langle{\cal O}^\psi_8({}^1S_0) \big\rangle
        +25.3  {\big\langle{\cal O}^\psi_8({}^3P_0) \big\rangle
       \over m^2_c} \in [0.24,0.45]~{\rm GeV}^3  \; ,
\end{equation}
where the range we have chosen takes into account theoretical
uncertainties due to $v^2$ relativistic corrections as mentioned previously.
Note that the above constraints are sufficient to insure that the helicity
rates in
Eq.~\ref{Rlambda} are positive for each value of the helicity
$\lambda$.

To the above experimental constraints on the allowed parameter space,
we must of course add the theoretical requirement that the
absolute values of the octet matrix elements
respect $v$-scaling rules, {\it i.e.}
that they must not be unreasonably larger than
$v^4\langle{\cal O}^\psi_1({}^3S_1) \rangle \sim 0.1$.
Moreover,
we must impose on our allowed volume of parameter space
that the matrix element
$\langle{\cal O}^\psi_8({}^1S_0) \rangle$
always be positive.  No similar constraint need
to be applied to the matrix element
$\langle{\cal O}^\psi_8({}^3P_0) \rangle$.  Indeed,
it can be negative, for the following reason.
The bare matrix element (which is
of course positive definite) contains a power ultraviolet divergence.
Such a power ultraviolet divergence must be proportional to a matrix
element of an operator of lower dimension.  In the case of
$\langle{\cal O}^\psi_8({}^3P_0) \rangle/m^2_c$, the divergence is
proportional to
$\langle {\cal O}^\psi_1({}^3S_1) \rangle $ \cite{bbl}.
This divergence must be subtracted to
obtain the renormalized matrix element.  Since the
piece subtracted is comparable in magnitude to the bare
matrix element, the difference can be negative.
On the other hand, since the operator
${\cal O}^\psi_8({}^1S_0)$ belongs to the set of
lowest dimension NRQCD four-fermion operators, the bare matrix element
$\langle{\cal O}^\psi_8({}^1S_0) \rangle$ cannot have power
divergences.
Thus the subtractions
involved in renormalization cannot transform the
positive definite bare matrix
element into a negative quantity~\cite{ebpc}.
Due to these considerations, we therefore impose the
additional theoretical constraint
\begin{equation}
\label{s-constraint}
\langle {\cal O}^\psi_8({}^1S_0) \rangle  > 0.
\end{equation}

The constraints expressed in eqs.~\ref{clm2}, \ref{afm2},
\ref{decayconstraint2} and \ref{s-constraint}
together define the volume
in the three-dimensional parameter space that is allowed
by current experimental information and theoretical considerations.

In defining the allowed parameter space,
we have not used an experimental constraint on the CDF-related linear
combination presented in Eq.~\ref{clspconstraint}.
Interestingly, we find that with the imposition of
all the above constraints
(Eqs.~\ref{clm2} through \ref{s-constraint}),
this combination
is limited to take values in the range
\begin{equation}
\label{clps2}
\big\langle {\cal O}^\psi_8(^1S_0) \big\rangle + 3
{\big\langle {\cal O}^\psi_8(^3P_0) \big\rangle\over m^2_c}
\in [0.01, 0.06]~{\rm GeV}^3 \; .
\end{equation}
The central value of 0.066 quoted in Eq.~\ref{clspconstraint}
falls just outside the upper end of this range,
indicating a possible inconsistency between
the result of Ref.~\cite{cl} and the results {}from other
processes considered in the present analysis.

We now present our main result, which is the expected range of $\alpha$:
The {\it maximum} value for $\alpha$ ($-0.09$) is obtained when
$\langle{\cal O}^{\psi}_8(^3S_1)\rangle$ is at the minimum
of its allowed range and when the photoproduction-related combination
$\langle {\cal O}^{\psi}_8(^1S_0)\rangle + 7
\langle {\cal O}^{\psi}_8(^3P_0)\rangle/m_c^2$ is at the maximum
of its allowed range. The {\it minimum}
value for $\alpha$ ($-0.33$) occurs in the opposite situation.

So far in our discussion, we have considered only
the particular choice $m_b = 4.7$ GeV.
Table I shows how our findings depend
on $m_b$.
The main results are the NRQCD predictions for the range
of $\alpha$.
Alongside the full NRQCD results, we
display the color-singlet model predictions
for $\alpha$.
The error bars on the color-singlet-model predictions
reflect $v^4$ relativistic
corrections; the relativistic corrections to the color-singlet result
are of order $v^4$, not $v^2$ as one might expect {\it a priori},
because
the $v^2$ corrections to the helicity
rates factor  and cancel in the ratio $\alpha$.
In the other three columns of Table I,
we treat the quantities appearing at the top of the
columns as functions of the variables through which we scan;
we report the ranges of values of these
quantities that occur for the allowed parameter space.

\begin{table}
$$\vbox{\tabskip=0pt \offinterlineskip
\halign to \hsize{\strut#& \vrule#\tabskip 1em plus 2em minus .5em&
\hfil#\hfil &\vrule#& \hfil#\hfil &\vrule#& \hfil#\hfil &\vrule#&
\hfil#\hfil &\vrule#& \hfil#\hfil &\vrule#& \hfil#\hfil &\vrule#
\tabskip=0pt\cr
\noalign{\hrule}
&& && && && && && &\cr
&& $m_b$ && color-singlet && NRQCD
&&range of && range of && range of &\cr
&& && model predictions && factorization formalism &&
$\big\langle{\cal O}^{\psi}_8(^1S_0)\big\rangle + 3
{\big\langle{\cal O}^{\psi}_8(^3P_0)\big\rangle \over m^2_c}$
&& $\big\langle{\cal O}^{\psi}_8(^1S_0)\big\rangle$ &&
${\big\langle{\cal O}^{\psi}_8(^3P_0)\big\rangle \over m^2_c}$ &\cr
&& (GeV) && for $\alpha$ && predictions for $\alpha$
&&(GeV$^3$)&& (GeV$^3$)&& (GeV$^3$) &\cr
&& && && && && && &\cr
\noalign{\hrule}
&& && && && && && &\cr
&& 4.1 && $-.27\pm .03$ && [$-.23$, $-.14$] && [.08,.2] &&
[.1,.4] && [$-.06$,$-.009$] &\cr
&& && && && && && &\cr
\noalign{\hrule}
&& && && && && && &\cr
&& 4.4 && $-.34\pm .03$ && [$-.28$, $-.11$] && [.03,.1] &&
[.02,.2] && [$-.03$,.003]&\cr
&& && && && && && &\cr
\noalign{\hrule}
&& && && && && && &\cr
&& 4.7 && $-.40\pm .04$ && [$-.35$, $-.08$] && [.009,.06] &&
[$0$,.1] && [$-.01$,.006]&\cr
&& && && && && && &\cr
\noalign{\hrule}
&& && && && && && &\cr
&& 5.0 && $-.45\pm .05$ && [$-.42$, $-.09$] && [.002,.03] &&
[$0$,.06] && [$-.008$,.006]&\cr
&& && && && && && &\cr
\noalign{\hrule}
&& && && && && && &\cr
&& 5.3 && $-.49\pm .05$ && [$-.47$, $-.12$] && [.001,.02] &&
[$0$,.03] && [-.005,.005]&\cr
&& && && && && && &\cr
\noalign{\hrule}  }}$$
\caption{ Theoretical leading-order predictions for
the expected range of $\alpha$.
We present predictions for
$\alpha$ in the color-singlet model and in the NRQCD
factorization formalism.  The ranges are determined by
``scanning" through the  allowed volume in the
three-dimensional parameter space
constrained by Eqs. (12), (13), (14) and (15) and by finding
within that volume the
maximum and minimum occuring values of $\alpha$.
Results are given for various
values of $m_b$.
Also given, in the last three columns, are the
ranges of the CDF-related combination and
of the two matrix elements indicated,
resulting from
the set of constraints.
}
\end{table}

Our leading-order results depend strongly on the values taken for $m_b$ and
$m_c$.
As stated previously we have used $m_c = 1.55$~GeV to
generate the results given in Table~I. For
some optional choices of $m_c$ we
obtain
the following ranges of $\alpha$. Choosing $m_b =4.7$~GeV
and $m_c = 1.25$~GeV we find that $\alpha \in [-.53,-.12]$.
Choosing $m_b = 4.7$~GeV and $m_c = 1.85$~GeV we find
that $\alpha \in [-.22,-.13]$.

In general
for higher values of $m_b$ the predicted range for $\alpha$
enlarges. For increasing values of $m_c$ the minimum value we predict
for $\alpha$ becomes greater, while the maximum value
remains roughly unchanged.

A general conclusion that can be drawn by looking at
Table I is that the inclusion of octet matrix
elements {\it raises} the predicted range for $\alpha$ significantly
{}from the color-singlet prediction.

\section{Conclusion}

We have proposed that
a measurement of the polarization of $J/\psi$ particles produced in
the process $b \to J/\psi + X$
can furnish a useful new means of determining
the color-octet matrix elements involved in
$J/\psi$ production.
This measurement, which is not yet available,
will supplement existing experimental
information, which includes extractions
{}from CDF, photoproduction, and the helicity-summed
rate $\Gamma(b \to J/\psi + X)$.
Further experimental information would be of great utility, since
there currently prevails
a certain inconsistency between the values
of the matrix elements measured at CDF
and those values extracted {}from other processes.
%

We have presented the factorization formula
for the helicity-summed rate in Eq.~\ref{totalfacform}.
Rates for specified $J/\psi$ helicity have been calculated using the methods
of Beneke and Rothstein \cite{br} and Braaten and Chen \cite{bc},
and are presented in Eq.~\ref{Rlambda}.  The resulting
expression for the polarization parameter $\alpha$ is
given in Eq.~\ref{bigresult} (for the choice $m_b = 4.7~{\rm GeV}$).

In anticipation of the measurement of the polarization
parameter $\alpha$, we have found it
interesting to determine the range of $\alpha$ which
is predicted by the NRQCD factorization formalism, given
current experimental information on the NRQCD matrix elements.
We find that, for $m_b = 4.7~{\rm GeV}$,
$\alpha$ is expected to range {}from $-0.33$ to $-0.09$.
Our method of determining this range consists of
``scanning" through the three-dimensional parameter space
of the color-octet matrix elements
that is
allowed by the constraints
expressed in Eqs.~\ref{clm2}, \ref{afm2}, \ref{decayconstraint2}
and \ref{s-constraint}.
Upon inspecting the parameter-space volume so defined,
we find that the linear combination of matrix elements that
has been measured at CDF \cite{cl},
namely $\langle {\cal O}^\psi_8(^1S_0) \rangle + 3
\langle {\cal O}^\psi_8(^3P_0) \rangle/ m^2_c$
is limited to take values
{}from 0.01 to 0.06 GeV$^3$, a situation which
is at the threshold of incompatibility
with the value of $0.066 \pm 0.015$ reported in
Ref.~\cite{cl}.

In Table I, we have shown how our
leading order predictions depend on $m_b$.
In general, it can be concluded that
the inclusion of octet matrix
elements raises the predicted range for $\alpha$ significantly
{}from the color-singlet prediction, and
that the range of $\alpha$ broadens as $m_b$ is increased.

Unfortunately, due to the large number of poorly determined
parameters, including $m_b$ and $m_c$,  the expected
range of $\alpha$ as predicted by NRQCD is large.

It may be that an accurate NRQCD prediction of the polarization parameter
(beyond leading order)
will not be possible for a long time. This possibility, however, does
not degrade the importance of our calculation,
since an experimental
determination of $\alpha$ ---
in conjuction with our results --- will
most certainly serve to
tighten the constraints on the possible values of the color-octet matrix
elements. In fact, the NRQCD prediction for $\alpha$ is
very sensitive to the values of the color-octet matrix elements,
and this offers
the hope that a measurement of the polarization of $J/\psi$ produced
in $b \to J/\psi+X$ will be instrumental in determining
$\langle{\cal O}^\psi_8({}^3S_1) \rangle$,
$\langle{\cal O}^\psi_8({}^1S_0) \rangle$, and
$\langle{\cal O}^\psi_8({}^3P_0) \rangle$.

\bigskip\centerline{{\bf Acknowledgements}}

We would especially like to thank Eric Braaten for many helpful discussions,
and Geoffrey T. Bodwin for discussions
regarding the $b$-quark mass.  I.M. and S.F gratefully acknowledge
James Buck Monta\~no for his invaluable help with the computer
programming in various numerical analyses. We also wish to
acknowledge the hospitality of the University of Wisconsin-Madison,
Ohio State University (I.M.), and the University of Texas at Austin (S.F).
The work of S.F. was supported in
part by the U.S.~Department of Energy under Grant
no.~DE-FG02-95ER40896, in part by the University of Wisconsin Research
Committee with funds granted by the Wisconsin Alumni Research
Foundation.  The work of I.M. was supported by the Robert A. Welch
Foundation, by NSF Grant PHY 9511632, and by NSERC of Canada.  The
work of O.F.H. and H.N. was supported by les Fonds FCAR du Qu\'ebec and
by NSERC of Canada.


\vfill\eject


\centerline{   \bf Figure Captions}

\bigskip\bigskip

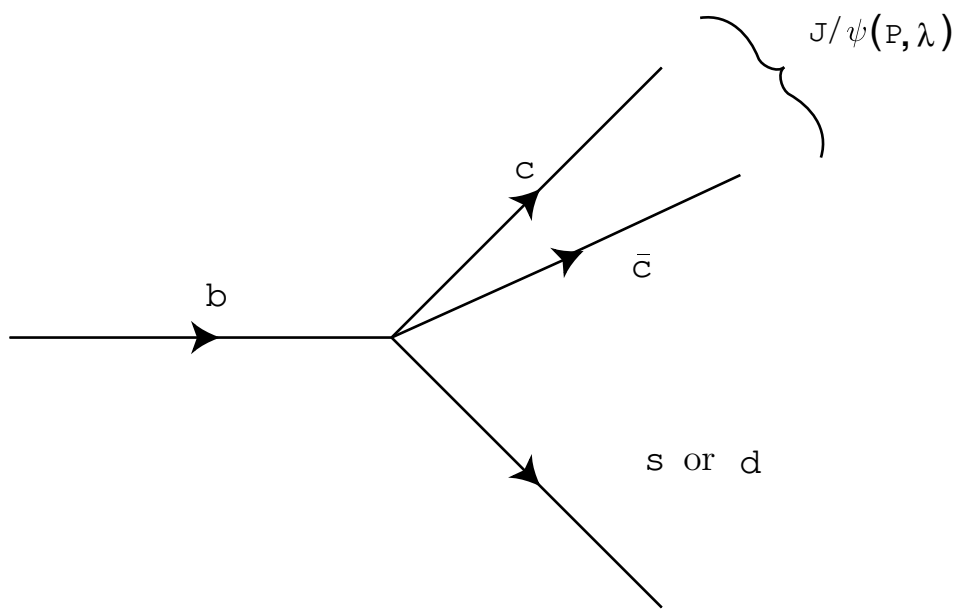
\noindent {\bf Figure 1}.  Feynman diagram for the four-quark
Fermi interactions $b \to c\bar{c}s$ and
$b \to c\bar{c}d$.  This interaction describes the short-distance
physics for the process $b \to J/\psi + X$.  $P$ is the total
four-momentum of the $c\bar{c}$ system, and $s$, $d$ and $b$ are
the four-momenta of the corresponding quarks.
The $c\bar{c}$ pair is produced initially with
quantum numbers $a,{}^{2S+1}L_J$ with $ a = 1 \, {\rm or} \, 8$.
It
then hadronizes into a $J/\psi$ particle with helicity $\lambda$.

\bigskip\bigskip

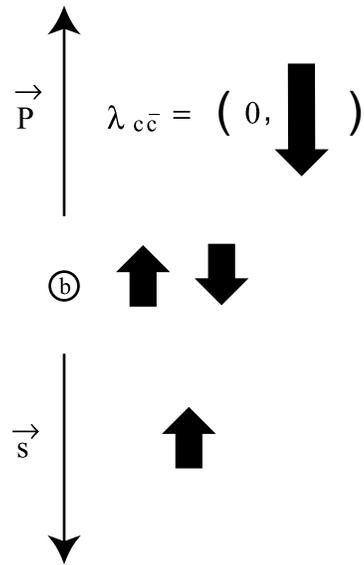
\noindent {\bf Figure 2}.  Angular momentum conservation in
 $b \to c\bar{c}s$ and
$b \to c\bar{c}d$.  Because the charged-current Fermi-interaction
couples only the left-handed parts of the fermion fields,
the $s$ and $d$ quarks (being massless) are emitted with negative helicity.
Due to angular momentum conservation, therefore,
the outcome $\lambda_{c\bar{c}}= +1$ is not allowed.
As a result, the
$\langle{\cal O}^\psi_1({}^3S_1) \rangle$
and $\langle{\cal O}^\psi_8({}^3S_1) \rangle$
contributions to the $J/\psi$ helicity rate are zero for $\lambda = +1$,
at lowest order in $v^2$.

\end{document}